\documentclass[12pt]{elsart}

\usepackage{amsmath,amssymb}
\usepackage{feynmp}
\usepackage[dvips]{graphicx}

\newcommand\GeV{\,\text{GeV}}

\newcounter{Section}
\newcommand\Section{%
  \stepcounter{Section}%
  \par\vspace{\baselineskip}\noindent\textbf{\arabic{Section}.}\quad
}

\journal{Physics Letters B}

\date{19 April 2006}

\begin{document}

\begin{frontmatter}

\title{Double Transverse-Spin Asymmetries in Drell--Yan Processes
with Antiprotons}

\author[al]{Vincenzo Barone},
\author[le,cr]{Alessandro Cafarella},
\author[le]{Claudio Corian\`{o}},
\author[le,co]{Marco Guzzi} and
\author[co]{Philip G.~Ratcliffe}

\address[al]{Di.S.T.A., Universit\`{a} del Piemonte Orientale ``A.~Avogadro'', \\
  and INFN, Gruppo Collegato di Alessandria, 15100 Alessandria, Italy}
\address[le]{Dipartimento di Fisica, Universit\`{a} di Lecce, \\
  and INFN, Sezione di Lecce, 73100 Lecce, Italy}
\address[cr]{Department of Physics, University of Crete,
  71003 Heraklion, Greece}
\address[co]{Dipartimento di Fisica e Matematica, Universit\`{a} dell'Insubria, \\
  22100 Como, Italy,  and INFN, Sezione di Milano, 20133 Milano, Italy}

\begin{abstract}
We present next-to-leading order predictions for double transverse-spin
asymmetries in Drell--Yan dilepton production initiated by proton--antiproton
scattering. The kinematic region of the proposed PAX experiment at GSI:
$30\lesssim{s}\lesssim200\GeV^2$ and $2\lesssim{M}\lesssim7\GeV$ is examined.
The Drell--Yan asymmetries turn out to be large, in the range 20--40\%.
Measuring these asymmetries would provide the cleanest
determination of the quark transversity distributions.
\end{abstract}


\end{frontmatter}

\newpage

\Section
The experiments with antiproton beams planned for the next decade in
the High-Energy Storage Ring at GSI will provide a variety of perturbative and
non-perturbative tests of QCD~\cite{brodsky}. In particular, the possible
availability of \emph{transversely polarised} antiprotons opens the way to
direct investigation of transversity, which is currently one of the main goals
of high-energy spin physics~\cite{bdr}. The quark transversity (\emph{i.e.}
transverse polarisation) distributions $\Delta_Tq$ were first introduced and
studied in the context of transversely polarised Drell--Yan (DY)
production~\cite{rs}; this is indeed the cleanest process probing these
quantities. In fact, whereas in semi-inclusive deep-inelastic scattering
transversity couples to another unknown quantity, the Collins fragmentation
function~\cite{collins}, rendering the extraction of $\Delta_Tq$ a not
straightforward task, the DY double-spin asymmetry
\begin{equation}
  A_{TT}^{DY}
  \equiv
  \frac{
    \d\sigma^{\uparrow \uparrow} - \d\sigma^{\uparrow \downarrow}
  }{
    \d\sigma^{\uparrow \uparrow} - \d\sigma^{\uparrow \downarrow}
  }
  =
  \frac{\Delta_T \sigma}{\sigma_\text{unp}}
  \label{att1}
\end{equation}
only contains combinations of transversity distributions. At leading order, for
instance, for the process $p^{\uparrow}p^{\uparrow}\to\ell^+\ell^-X$ one has
\begin{equation}
  A_{TT}^{DY} =
  a_{TT} \,
  \frac{\sum_q e_q^2
    [
      \Delta_T      q (x_1, M^2) \, \Delta_T \bar{q}(x_2, M^2) +
      \Delta_T \bar{q}(x_1, M^2) \, \Delta_T      q (x_2, M^2)
    ]
  }{\sum_q e_q^2
    [q(x_1, M^2) \, \bar{q}(x_2, M^2) + \bar{q}(x_1, M^2) \, q(x_2, M^2)]
  },
  \label{att2}
\end{equation}
where $M$ is the invariant mass of the lepton pair, $q(x,M^2)$ is the
unpolarised distribution function, and $a_{TT}$ is the spin asymmetry of the QED
elementary process $q\bar{q}\to\ell^+\ell^-$. In the dilepton centre-of-mass
frame, integrating over the production angle $\theta$, one has
\begin{equation}
  a_{TT} (\varphi) = \half \, \cos 2 \varphi \,,
\end{equation}
where $\varphi$ is the angle between the dilepton direction and the plane
defined by the collision and polarisation axes.

Measurement of $p^{\uparrow}p^{\uparrow}$ DY is planned at RHIC~\cite{rhic}. It
turns out, however, that $A_{TT}^{DY}(pp)$ is rather small at such
energies~\cite{bcd,mssv,ratcliffe}, no more than a few percent (similar values
are found for double transverse-spin asymmetries in prompt-photon
production~\cite{Mukherjee:2003pf} and single-inclusive hadron
production~\cite{Mukherjee:2005rw}). The reason is twofold: 1) $A_{TT}^{DY}(pp)$
depends on antiquark transversity distributions, which are most likely to be
smaller than valence transversity distributions; 2) RHIC kinematics
($\sqrt{s}=200\GeV$, $M<10\GeV$ and $x_1x_2=M^2/s\lesssim3\times10^{-3}$) probes
the low-$x$ region, where QCD evolution suppresses $\Delta_Tq(x,M^2)$ as
compared to the unpolarised distribution $q(x,M^2)$~\cite{bcd2,Barone:1997fh}.
The problem may be circumvented by studying transversely polarised
proton--\emph{anti}proton DY production at more moderate energies. In this case
a much larger asymmetry is expected~\cite{bcd,Anselmino:2004ki,Efremov:2004qs}
since $A_{TT}^{DY}(p\bar{p})$ is dominated by valence distributions at medium
$x$. The PAX collaboration has proposed the study of
$p^{\uparrow}\bar{p}^{\uparrow}$ Drell--Yan production in the High-Energy
Storage Ring (HESR) at GSI, in the kinematic region
$30\GeV^2\lesssim{s}\lesssim200\GeV^2$, $2\GeV\lesssim{M}\lesssim10\GeV$ and
$x_1x_2\gtrsim0.1$~\cite{pax}. An antiproton polariser for the PAX experiment is
currently under study~\cite{pax_prl}: the aim is to achieve a polarisation of
30--40\%, which would render the measurement of $A_{TT}^{DY}(p\bar{p})$ very
promising.

Leading-order predictions for the $p\bar{p}$ asymmetry at moderate $s$ were
presented in~\cite{Anselmino:2004ki}. It was also suggested there to access
transversity in the $J/\psi$ resonance production region, where the production
rate is much higher. The purpose of this paper is to extend the calculations
of~\cite{Anselmino:2004ki} to next-to-leading order (NLO) in QCD.\footnote{The
results presented here were communicated at the QCD--PAC meeting at GSI (March
2005) and reported by one of us (M.G.) at the Int. Workshop ``Transversity 2005''
(Como, September 2005)~\cite{guzzi}.} This is a necessary check of the previous
conclusions, given the moderate values of $s$ in which we are interested. We
shall see that the NLO corrections are actually rather small and double
transverse-spin asymmetries are confirmed to be of order 20--40\%.

\begin{figure}[hbt]
  \centering
  \begin{fmffile}{attnlo-fmf}
    \setlength\unitlength{1.0mm}
    \fmfset{arrow_len}{2.5mm}
    \fmfset{curly_len}{3.0mm}
    \fmfset{wiggly_slope}{75}
    \begin{fmfgraph}(30,30)
      \fmfleft{i2,i1}
      \fmfright{o1}
      \fmf{fermion,tension=2.0}{i1,v1,v2,v3,i2}
      \fmf{photon,tension=1.5}{v2,o1}
      \fmffreeze
      \fmf{gluon,right=0.3}{v1,v3}
    \end{fmfgraph}%
    \qquad
    \begin{fmfgraph}(30,30)
      \fmfset{curly_len}{2.0mm}
      \fmfleft{i2,i1}
      \fmfright{o1}
      \fmf{fermion,tension=3.0}{i1,v1,v2,v3}
      \fmf{fermion}{v3,i2}
      \fmf{photon,tension=1.5}{v3,o1}
      \fmffreeze
      \fmf{gluon,right}{v2,v1}
    \end{fmfgraph}%
    \qquad
    \fmfframe(0,1)(0,1){%
    \begin{fmfgraph}(50,28)
      \fmfleft{i2,i1}
      \fmfright{o2,o1}
      \fmf{fermion}{i1,v1}
      \fmf{fermion,l.s=right,tension=0.0}{v1,v2}
      \fmf{fermion}{v2,i2}
      \fmf{photon}{v1,o1}
      \fmf{gluon}{o2,v2}
    \end{fmfgraph}%
    }%
  \end{fmffile}%
  \\[1ex]
  (a)\hspace{33mm}(b)\hspace{44mm}(c)\hspace{10mm}
  \caption{Elementary processes contributing to the transverse Drell--Yan
           cross-section at NLO: (a, b) virtual-gluon corrections and (c)
           real-gluon emission.}
  \label{dyqqg2}
\end{figure}
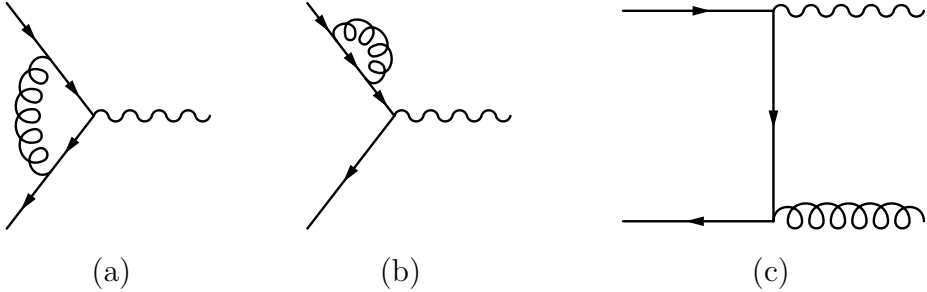

\Section
The kinematic variables describing the Drell--Yan process are (1 and 2
denote the colliding hadrons):
\begin{equation}
  \xi_1 = \sqrt{\tau} \, e^y \,, \qquad
  \xi_2 = \sqrt{\tau} \, e^{-y} \,, \qquad
  y = \frac{1}{2} \, \ln \frac{\xi_1}{\xi_2} \,,
  \label{kin}
\end{equation}
with $\tau=M^2/s$. We denote by $x_1$ and $x_2$ the longitudinal momentum
fractions of the incident partons. At leading order, $\xi_1$ and $\xi_2$
coincide with $x_1$ and $x_2$, respectively. The QCD factorisation formula for
the transversely polarised cross-section for the proton--antiproton Drell--Yan
process is
\begin{multline}
  \frac{\d\Delta_T \sigma}{\d M \, \d y \,\d\varphi}
  =
  \sum_q e_q^2 \int_{\xi_1}^1 \d x_1 \int_{\xi_2}^1 \d x_2
  \left[ \Delta_T q(x_1,\mu^2) \Delta_T q(x_2,\mu^2) \right.
\\
  \null +
  \left. \Delta_T \bar{q}(x_1,\mu^2) \Delta_T \bar{q}(x_2,\mu^2) \right]
  \frac{\d\Delta_T \hat\sigma}{\d M \, \d y \, \d\varphi}
  \,,
  \label{fact}
\end{multline}
where $\mu$ is the factorisation scale
and we take the quark (antiquark) distributions of the antiproton equal to the
antiquark (quark) distributions of the proton.
Note that, since gluons cannot be transversely polarised (there is no such thing
as a gluon transversity distribution for a spin one-half object like the
proton), only quark--antiquark annihilation subprocesses (with their radiative
corrections) contribute to $\d\Delta_T\sigma$. In Eq.~\eqref{fact} we use the
fact that antiquark distributions in antiprotons equal quark distributions in
protons, and \emph{viceversa}. At NLO, \emph{i.e.} at order $\alpha_s$, the
hard-scattering cross-section $\d\Delta_T\hat\sigma^{(1)}$, taking the diagrams
of Fig.~\ref{dyqqg2} into account, is given by~\cite{mssv}
\begin{align}
  \frac{
    \d\Delta_T \hat\sigma^{(1),\overline{\text{MS}}}
  }{
    \d M \, \d y \, \d\varphi
  }
  \hspace{-4em} & \hspace{4em} =
  \frac{2\alpha^2}{9 s M} \, C_F \,
  \frac{\alpha_s(\mu^2)}{2\pi} \,
  \frac{4\tau(x_1x_2+\tau)}{x_1x_2(x_1+\xi_1)(x_2+\xi_2)} \,
  \cos(2\varphi)
  \nonumber
  \\[1ex]
  & \hspace{-1em} \null \times
  \Bigg\{
  \delta(x_1-\xi_1)\delta(x_2-\xi_2) \! \left[
  \frac{1}{4}\ln^2\frac{(1-\xi_1)(1-\xi_2)}{\tau}
  +\frac{\pi^2}{4}-2 \right]
  \nonumber
  \\[1ex]
  & \null +
  \delta(x_1-\xi_1)
  \left[
    \frac{1}{(x_2-\xi_2)_{+}} \ln\frac{2x_2(1-\xi_1)}{\tau(x_2+\xi_2)}
    + \left(\frac{\ln(x_2-\xi_2)}{x_2-\xi_2}\right)_{\!+}
    + \frac{1}{x_2-\xi_2}\ln\frac{\xi_2}{x_2}
  \right]
  \nonumber
  \\[1ex]
  & \null
  + \frac{1}{2[(x_1-\xi_1)(x_2-\xi_2)]_{+}}
  + \frac{(x_1+\xi_1)(x_2+\xi_2)}{(x_1 \xi_2+x_2\xi_1)^2}
  - \frac{3\ln\!\left(\frac{x_1x_2+\tau}{x_1\xi_2+x_2\xi_1} \right)}
         {(x_1-\xi_1)(x_2-\xi_2)}
  \nonumber
  \\[1ex]
  & \null +
  \ln \frac{M^2}{\mu^2} \left[
  \delta(x_1-\xi_1)\delta(x_2-\xi_2) \!
  \left( \frac{3}{4}+\frac{1}{2}\ln \frac{(1-\xi_1)(1-\xi_2)}{\tau} \right)
  +
  \frac{\delta(x_1-\xi_1)}{(x_2-\xi_2)_{+}} \right] \! \Bigg\}
  \nonumber
  \\[1ex]
  & \hspace{25em} \null + \big[ 1 \leftrightarrow 2 \big],
  \label{sigmanlo}
\end{align}
where we have taken the factorisation scale $\mu$ equal to the renormalisation
scale. In our calculations we set $\mu = M$.

The unpolarised Drell--Yan differential cross-section can be found, for
instance, in~\cite{sutton}; besides the diagrams of Fig.~\ref{dyqqg2}, it also
includes the contribution of quark--gluon scattering processes.

\Section
To compute the Drell--Yan asymmetries we need an assumption for the
transversity distributions, which as yet are completely unknown. We might
suppose, for instance, that transversity equals helicity at some low scale, as
suggested by confinement models~\cite{bcd2} (this is exactly true in the
non-relativistic limit). Thus, one possibility is
\begin{equation}
  \Delta_T q(x,\mu_0^2) = \Delta q(x,\mu_0^2)\,,
  \label{minbound}
\end{equation}
where typically $\mu_0\lesssim1\GeV$. Another possible assumption for
$\Delta_Tq$ is the saturation of Soffer's inequality~\cite{soffer}, namely
\begin{equation}
  \big| \Delta_T q(x,\mu_0^2) \big| =
  \half \big[ q(x,\mu_0^2) + \Delta q(x,\mu_0^2) \big] ,
  \label{sofbound}
\end{equation}
which represents an upper bound on the transversity distributions.

Since Eqs.~\eqref{minbound} and \eqref{sofbound} make sense only at very low
scales, in practical calculations one has to resort to radiatively generated
helicity and number densities, such as those provided by the GRV
fits~\cite{grv}. The GRV starting scale is indeed (at NLO) $\mu_0^2=0.40\GeV^2$.
We should however bear in mind that in the GRV parametrisation there is a
sizeable gluon contribution to the nucleon's helicity already at the input scale
($\Delta g$ is of order $0.5$). On the other hand, as already mentioned, gluons
do not contribute to the nucleon's transversity. Thus, use of
Eq.~\eqref{minbound} with the GRV parametrisation may lead to substantially
underestimating the quark transversity distributions and hence is a sort of
``minimal bound'' for transversity. Incidentally, the experimental verification or
otherwise of the theoretical predictions of $A_{TT}$ based on the low-scale
constraints (\ref{minbound}, \ref{sofbound}) would represent an indirect test of
the ``valence glue'' hypothesis behind the GRV fits.
Note too that, although the assumption (\ref{minbound}) may, in principle,
violate the Soffer inequality, we have explicitly checked that this is not the
case with all the distributions we use.

\begin{figure}[hbt]
  \centering
  \includegraphics[width=7cm,angle=-90]{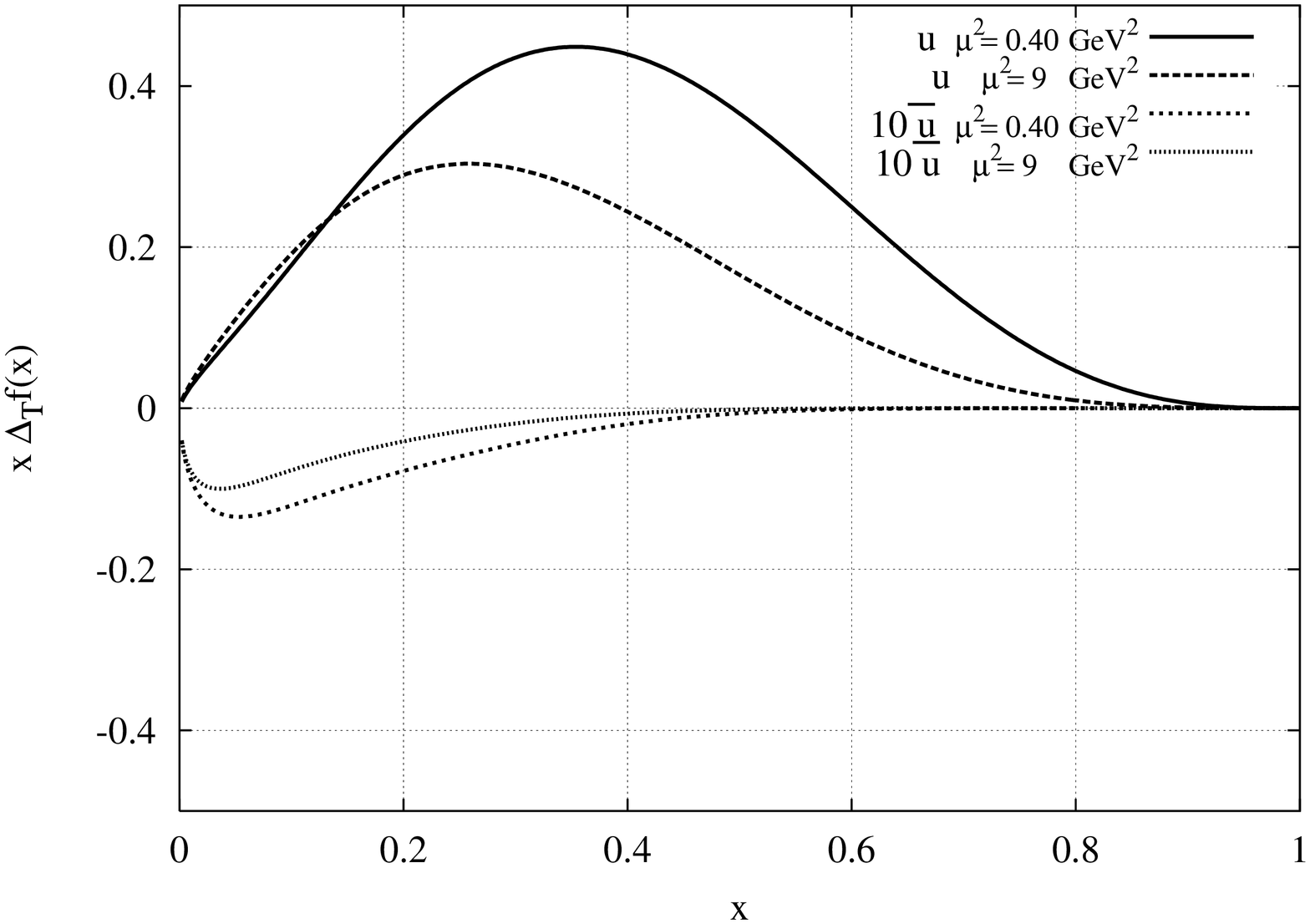}
  \includegraphics[width=7cm,angle=-90]{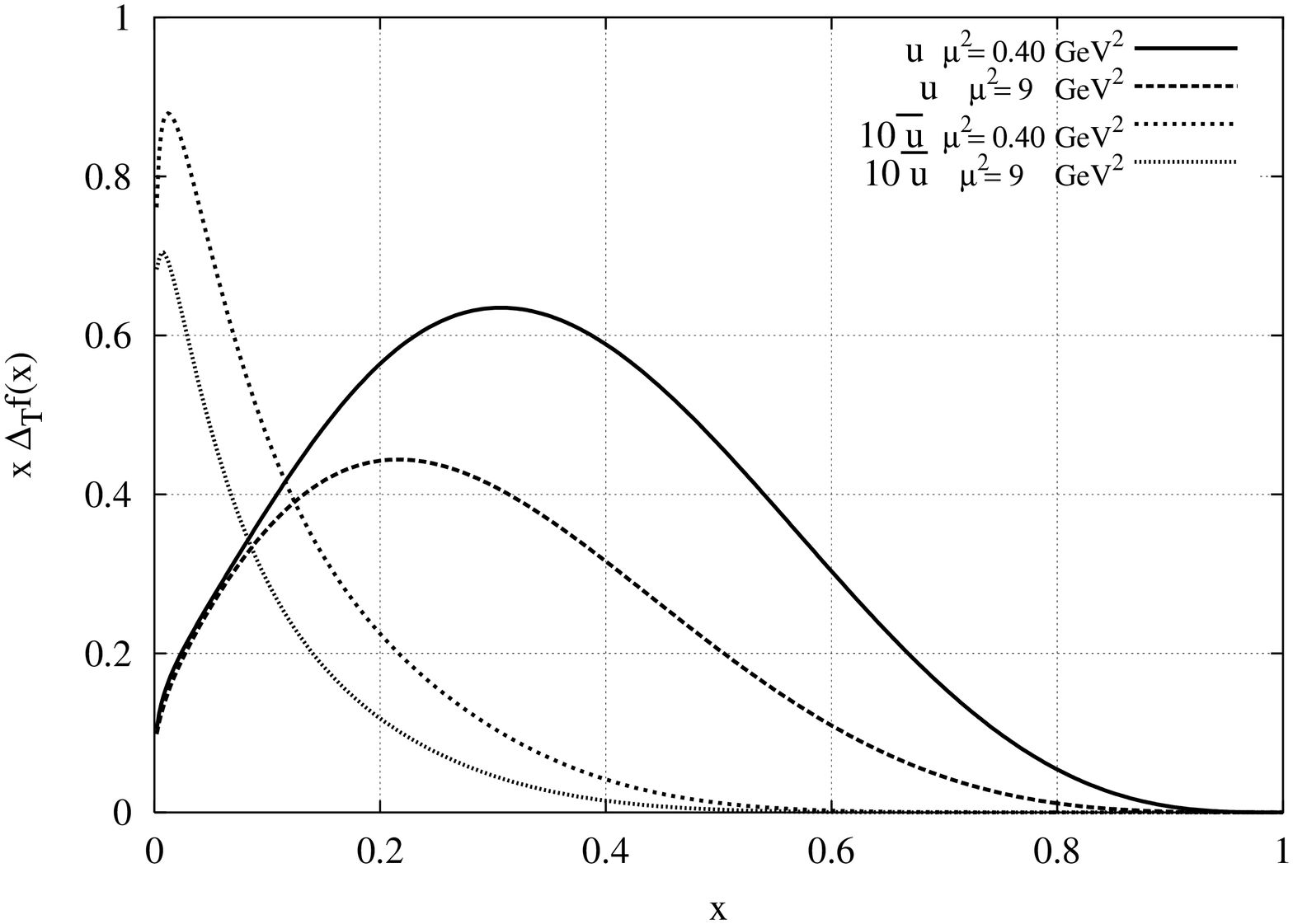}
  \caption{The $u$ and $\bar{u}$ transversity distributions, as obtained from
  the GRV parametrisation and Eq.~\eqref{minbound}, top panel, or
  Eq.~\eqref{sofbound}, bottom:
  $x\Delta_Tu$ at $\mu^2=\mu_0^2=0.40\GeV^2$ (dashed curve) and
  $\mu^2=9\GeV^2$ (solid curve);
  $x\Delta_T\bar{u}$ at $\mu^2=\mu_0^2=0.40\GeV^2$ (dotted curve) and
  $\mu^2=9\GeV^2$ (dot--dashed curve).
  Note that the $\bar{u}$ transversity distributions have been multiplied by a
  factor of $10$.}
  \label{umin}
\end{figure}

After setting the initial condition \eqref{minbound} or \eqref{sofbound}, all
distributions are evolved at NLO according to the appropriate DGLAP equations
(for transversity, see~\cite{h1evol}; the numerical codes we use to solve the
DGLAP equations are described in~\cite{cafarella}). The $u$ sector of
transversity is displayed in Fig.~\ref{umin} for the minimal bound
\eqref{minbound} and for the Soffer bound~\eqref{sofbound}.
%

\begin{figure}[hbt]
  \centering
  \includegraphics[width=7cm,angle=-90]{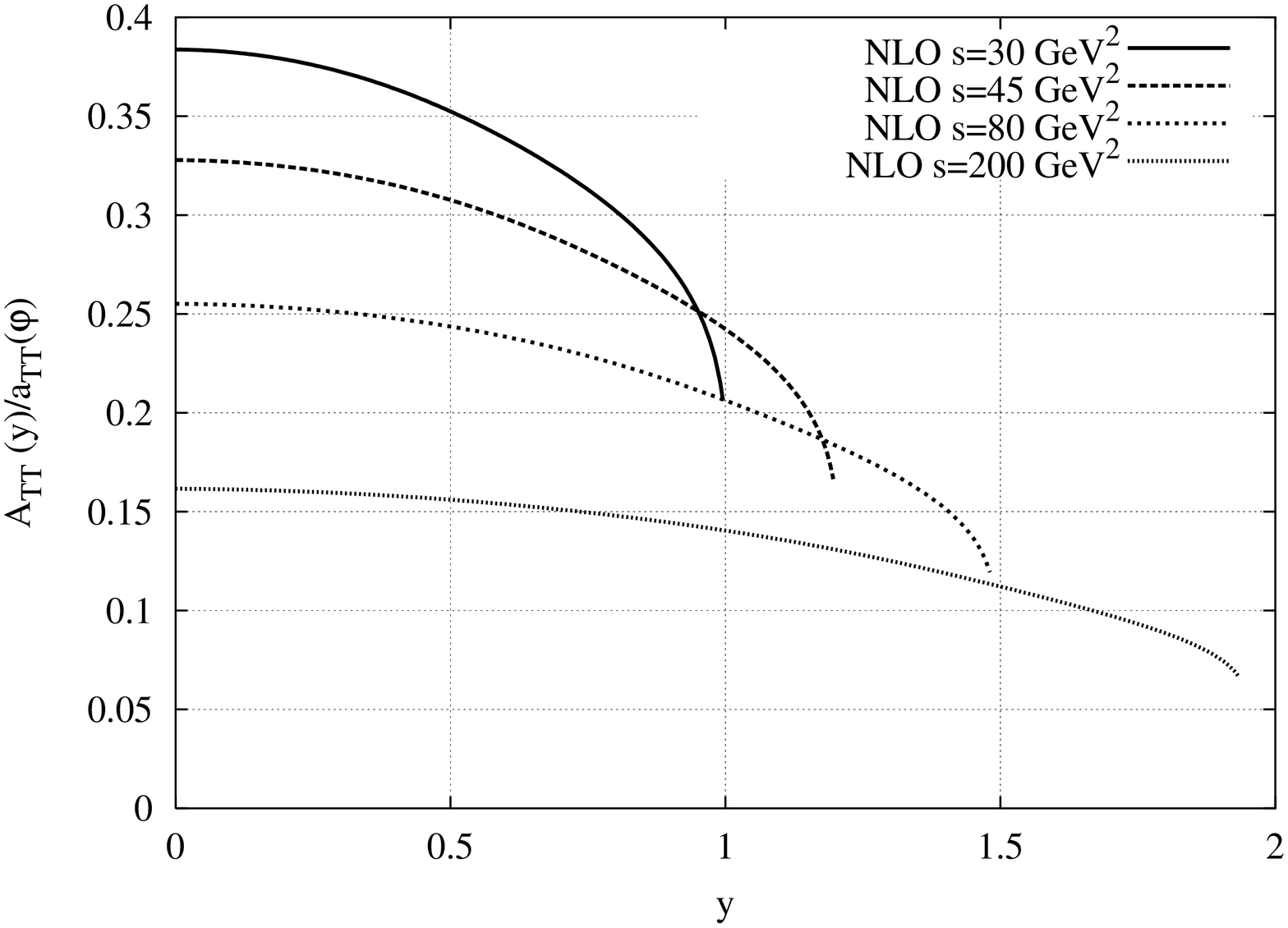}
  \caption{The NLO double transverse-spin asymmetry $A_{TT}(y)/a_{TT}$,
  integrated between $M=2\GeV$ and $M=3\GeV$, for various values of $s$; the
  minimal bound \eqref{minbound} is used for the input distributions.}
  \label{nlom23}
\end{figure}

\begin{figure}[hbt]
  \centering
  \includegraphics[width=7cm,angle=-90]{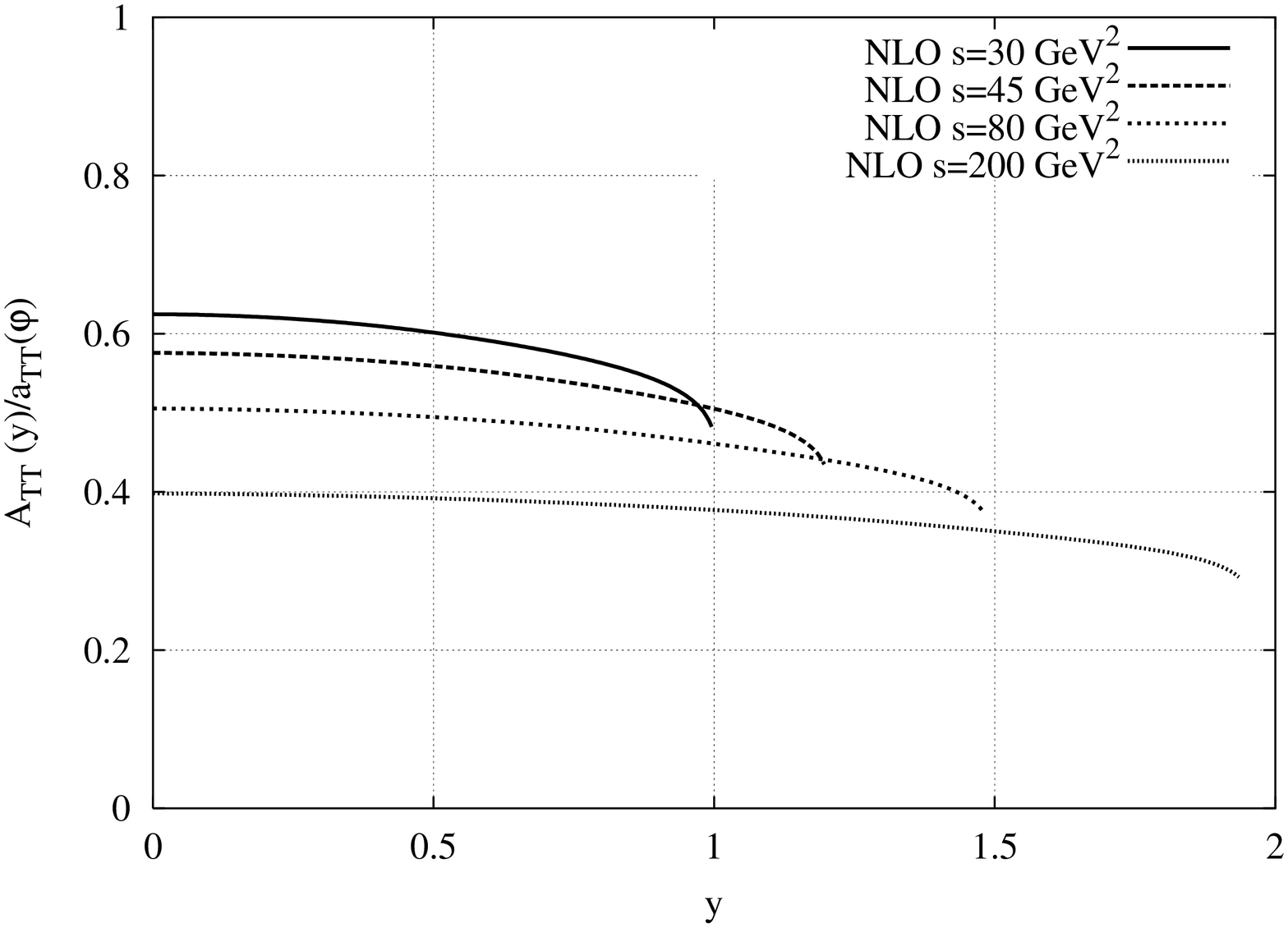}
  \caption{As Fig.~\ref{nlom23}, but with input distributions corresponding to
  the Soffer bound~\eqref{sofbound}.}
  \label{nlos23}
\end{figure}

The transverse Drell--Yan asymmetry $A_{TT}^{DY}/a_{TT}$, integrated over $M$
between $2\GeV$ and $3\GeV$ (\emph{i.e.} below the $J/\psi$ resonance region),
for various values of $s$ is shown in Fig.~\ref{nlom23}. As can be seen, the
asymmetry is of order of 30\% for $s=30\GeV^2$ (fixed-target option) and
decreases by a factor two for a centre-of-mass energy typical of the collider
mode ($s=200\GeV^2$). The corresponding asymmetry obtained by saturating the
Soffer bound, that is by using Eq.~\eqref{sofbound} for the input distributions,
is displayed in Fig.~\ref{nlos23}. As expected, it is systematically larger,
rising to over 50\% for fixed-target kinematics.

\begin{figure}[hbt]
  \centering
  \includegraphics[width=7cm,angle=-90]{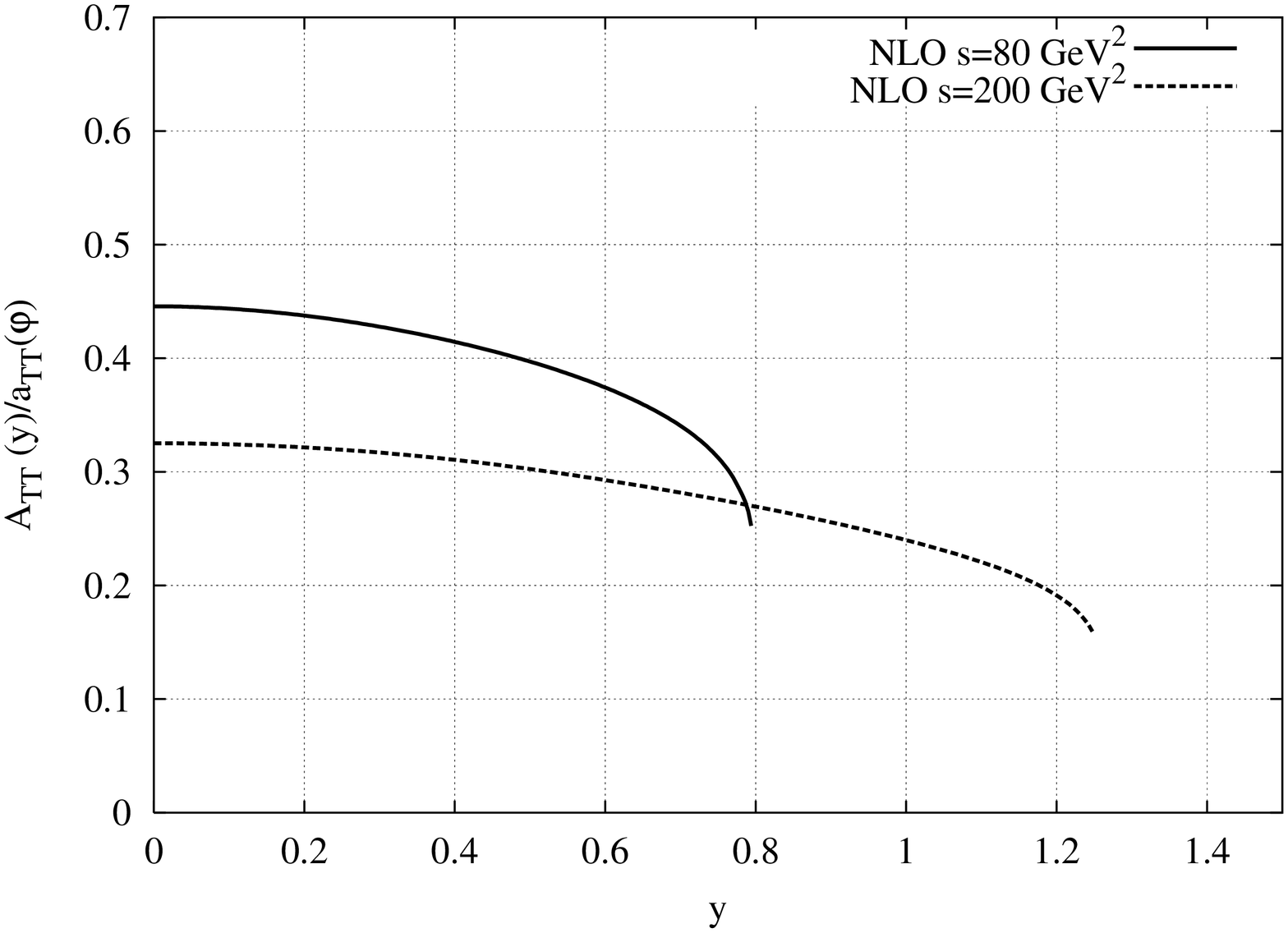}
  \caption{The NLO double transverse-spin asymmetry $A_{TT}(y)/a_{TT}$,
  integrated between $M=4\GeV$ and $M=7\GeV$, for various values of $s$; the
  minimal bound \eqref{minbound} is used for the input distributions.}
  \label{nlom47}
\end{figure}

Above the $J/\psi$ peak $A_{TT}^{DY}/a_{TT}$ appears as shown in
Fig.~\ref{nlom47}, where we present the results obtained with the minimal bound
\eqref{minbound}. Comparing Figs.~\ref{nlom23} and \ref{nlom47}, we see that the
asymmetry increases at larger $M$ (recall though that the cross-section falls
rapidly with growing $M$).

\begin{figure}[hbt]
  \centering
  \includegraphics[width=7cm,angle=-90]{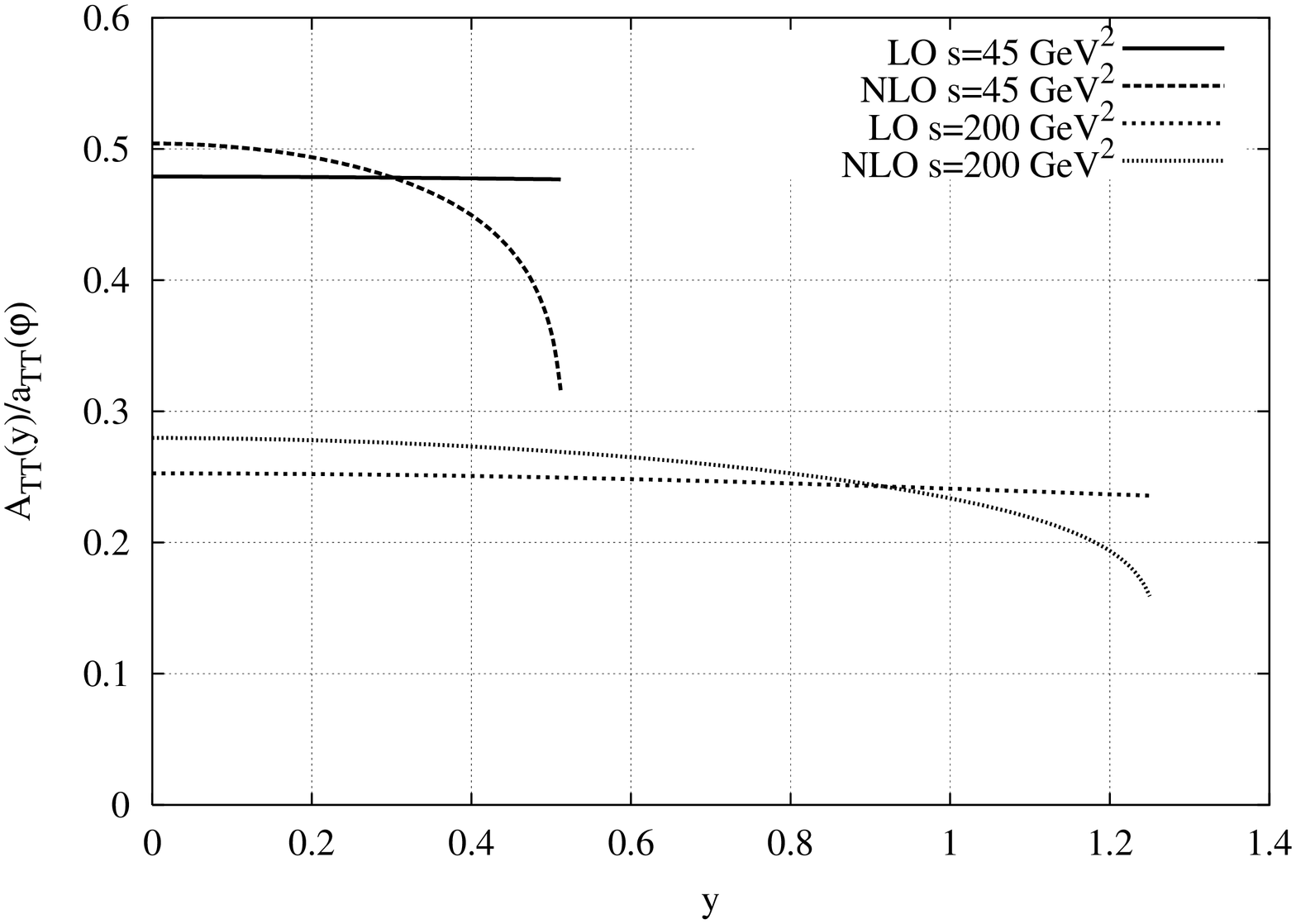}
  \caption{NLO \emph{vs.} LO double transverse-spin asymmetry $A_{TT}(y)/a_{TT}$
  at $M=4\GeV$ for $s=45\GeV^2$ and $s=200\GeV^2$; the minimal bound
  \eqref{minbound} is used for the input distributions.}
  \label{nlolo4}
\end{figure}

The importance of NLO QCD corrections may be appreciated from Fig.~\ref{nlolo4},
where one sees that the NLO effects hardly modify the asymmetry since the $K$
factors of the transversely polarised and unpolarised cross-sections are similar
to each other and therefore cancel out in the ratio.
As for the dependence on the factorisation scale $\mu$ (we recall that the
results presented in all figures are obtained setting $\mu = M$), we have
repeated the calculations with two other choices ($\mu = 2 M$ and $\mu = M/2$)
and found no sensible differences.

A \emph{caveat} is in order at this point. The GSI kinematics is dominated by
the domain of large $\tau$ and large $z=\tau/x_1x_2$, where real-gluon emission
is suppressed and where there are powers of large logarithms of the form
$\ln(1-z)$, which need to be resummed to all orders in
$\alpha_s$~\cite{shimizu}. It turns out that the effects of threshold
resummation on the asymmetry $A_{TT}^{DY}$ in the regime we are considering,
although not irrelevant, are rather small (about 10\%) if somewhat dependent on
the infrared cutoff for soft-gluon emission.

The feasibility of the $A_{TT}$ measurement at GSI has been thoroughly
investigated by the PAX Collaboration (see App.~F of~\cite{pax}). In collider
mode, with a luminosity of $5 \cdot 10^{30}\,\text{cm}^{-2}\,\text{s}^{-1}$, a
proton polarisation of 80\%, an antiproton polarisation of 30\% and considering
dimuon invariant masses down to $M = 2\GeV$, after one year's data taking one
expects a few hundred events per day and a statistical accuracy on $A_{TT}$ of
10--20\%.

\Section
Before concluding, we briefly comment on the possibility
of accessing transversity via $J/\psi$ production in $p\bar{p}$ scattering.
It is known that the dilepton production rate around $M=3\GeV$,
\emph{i.e.} at the $J/\psi$ peak, is two orders of magnitude higher than in the
region $M\simeq4\GeV$. Thus, with a luminosity of $5 \cdot
10^{30}\,\text{cm}^{-2}\,\text{s}^{-1}$, one expects a number of
$p\bar{p}\to{J}/\psi\to\ell^-\ell^+$ events of order $10^5$ per year at GSI
collider energies.
This renders the measurement of $A_{TT}$
in the $J/\psi$-resonance region extremely advantageous from a statistical
point of view.

As explained in~\cite{Anselmino:2004ki}, if $J/\psi$
formation is dominated by the $q\bar{q}$ annihilation channel,
at leading order the double
transverse-spin asymmetry at the $J/\psi$ peak has the same structure as the
asymmetry for Drell--Yan continuum production, since the $J/\psi$ is a vector
particle and the $q\bar{q}\,J/\psi$
coupling has the same helicity structure as
the $q\bar{q}\gamma^*$ coupling. The CERN SPS data~\cite{sps} show that the
$p\bar{p}$ cross-section for $J/\psi$ production at $s=80\GeV^2$ is about ten
times larger than the corresponding $pp$ cross-section, which is a strong
indication that the $q\bar{q}$-fusion mechanism is indeed dominant. Therefore,
at the $s$ values of interest here ($s\lesssim200\GeV^2$)
dilepton production in
the $J/\psi$ resonance region can be described in a manner analogous to
Drell--Yan continuum production, with the elementary subprocess
$q\bar{q}\to\gamma^*\to\ell^-\ell^+$ replaced by $q\bar{q}\to
J/\psi\to\ell^-\ell^+$~\cite{carlson}.
Using this model, which successfully accounts
for the SPS $J/\psi$ production data at moderate values of~$s$,
it was found in \cite{Anselmino:2004ki} that the transverse asymmetry
at the $J/\psi$ peak is of the order of 25--30\%.

At next-to-leading order,
due to QCD radiative corrections,
one cannot use a point-like $q\bar{q}\,J/\psi$
coupling, and therefore
it is not possible to extend in a straightforward way the
model used to evaluate $A_{TT}$ at leading order. Were NLO effects
not dominant, as is the case for continuum production,
one could still expect the $J/\psi$ asymmetry to be quite sizeable,
but this is no more than an educated guess.
What we wish to emphasise, however, is the
importance of experimentally investigating
the $J/\psi$ double transverse-spin
asymmetry, which can
shed light both on the transversity content of the
nucleon and on the mechanism of $J/\psi$ formation
(since gluon-initiated hard processes do not contribute
to the transversely polarised scattering,
the study of $A_{TT}$
in the $J/\psi$ resonance region may give information
on the relative weight of gluon and quark-antiquark subprocesses
in $J/\psi$ production).


\Section
In conclusion, experiments with polarised antiprotons at GSI will
represent a unique opportunity to investigate the transverse polarisation
structure of hadrons. The present paper, which confirms the results
of~\cite{Anselmino:2004ki}, shows that the double transverse-spin asymmetries
are large enough to be experimentally measured and therefore
represent the most promising observables to directly access
the quark transversity distributions.


\begin{ack}
We would like to thank M.~Anselmino, N.N.~Nikolaev and our colleagues of the PAX
collaboration for prompting the study reported here and for various useful
discussions. This work is supported in part by the Italian Ministry of
Education, University and Research (PRIN~2003).
\end{ack}


\begin{thebibliography}{99}

\bibitem{brodsky}
  S.J.~Brodsky, ``Testing Quantum Chromodynamics with Antiprotons'',
  hep-ph/0411046.

\bibitem{bdr}
  For a review on the transverse polarisation of quarks in hadrons, see
  V.~Barone, A.~Drago and P.G.~Ratcliffe,
  \emph{Phys. Rep.} \textbf{359} (2002)~1.

\bibitem{rs}
  J.~Ralston and D.E.~Soper, \emph{Nucl. Phys.} \textbf{B152} (1979)~109;
  J.L.~Cortes, B.~Pire and J.P.~Ralston, \emph{Z. Phys.} \textbf{C55}
  (1992)~409;
  R.L.~Jaffe and X.~Ji, \emph{Nucl. Phys.} \textbf{B375} (1992)~527.

\bibitem{collins}
  J.C.~Collins, \emph{Nucl. Phys.} \textbf{B396} (1993)~161.

\bibitem{rhic}
  G.~Bunce, N.~Saito, J.~Soffer and W.~Vogelsang,
  \emph{Ann. Rev. Nucl. Part. Phys.} \textbf{50} (2000)~525.

\bibitem{bcd}
  V.~Barone, T.~Calarco and A.~Drago, \emph{Phys. Rev.} \textbf{D56} (1997)~527.

\bibitem{mssv}
  O.~Martin, A.~Sch\"{a}fer, M.~Stratmann and W.~Vogelsang,
  \emph{Phys. Rev.} \textbf{D57} (1998) 3084; \emph{Phys. Rev.} \textbf{D60}
  (1999) 117502.

\bibitem{ratcliffe}
  P.G.~Ratcliffe, \emph{Eur. Phys. J.} \textbf{C41} (2005)~319.

\bibitem{Mukherjee:2003pf}
  A.~Mukherjee, M.~Stratmann and W.~Vogelsang, \emph{Phys. Rev.} \textbf{D67}
  (2003) 114006.

\bibitem{Mukherjee:2005rw}
  A.~Mukherjee, M.~Stratmann and W.~Vogelsang,
  \emph{Phys. Rev.} \textbf{D72} (2005) 034011.

\bibitem{bcd2}
  V.~Barone, T.~Calarco and A.~Drago, \emph{Phys. Lett.} \textbf{B390}
  (1997)~287.

\bibitem{Barone:1997fh}
  V.~Barone, \emph{Phys. Lett.} \textbf{B409} (1997)~499.

\bibitem{Anselmino:2004ki}
  M.~Anselmino, V.~Barone, A.~Drago and N.N.~Nikolaev,
  \emph{Phys. Lett.} \textbf{B594} (2004)~97.

\bibitem{Efremov:2004qs}
  A.V.~Efremov, K.~Goeke and P.~Schweitzer,
  \emph{Eur. Phys. J.} \textbf{C35} (2004)~207.

\bibitem{pax}
  $\mathcal{P\!AX}$ Collab.,
  V.~Barone \emph{et al.},
  ``Antiproton--Proton Scattering Experiments with Polarization'',
  Technical Proposal, hep-ex/0505054.

\bibitem{pax_prl}
  F.~Rathmann \emph{et al}., \emph{Phys. Rev. Lett.} \textbf{94} (2005) 014801.

\bibitem{guzzi}
  M.~Guzzi, in proc. of the Int. Workshop ``Transversity 2005'' (Como, 2005),
  V.~Barone and P.G.~Ratcliffe eds., World Scientific, in press.

\bibitem{sutton}
  P.J.~Sutton, A.D.~Martin, R.G.~Roberts and W.J.~Stirling,
  \emph{Phys. Rev.} \textbf{D45} (1992) 2349.

\bibitem{soffer}
  J.~Soffer, \emph{Phys. Rev. Lett.} \textbf{74} (1995) 1292.

\bibitem{grv}
  M.~Gl\"{u}ck, E.~Reya and A.~Vogt, \emph{Eur. Phys. J.} \textbf{C5} (1998)~461;
  M.~Gl\"{u}ck, E.~Reya, M.~Stratmann and W.~Vogelsang, \emph{Phys. Rev.}
  \textbf{D63} (2001) 094005.

\bibitem{h1evol}
  F.~Baldracchini, N.S.~Craigie, V.~Roberto and M.~Socolovsky, \emph{Fortschr.
  Phys.} \textbf{30} (1981) 505;
  X.~Artru and M.~Mekhfi, \emph{Z. Phys.} \textbf{C45} (1990) 669;
  A.~Hayashigaki, Y.~Kanazawa and Y.~Koike, \emph{Phys. Rev.} \textbf{D56}
  (1997) 7350;
  S.~Kumano and M.~Miyama, \emph{Phys. Rev.} \textbf{D56} (1997) R2504;
  W.~Vogelsang, \emph{Phys. Rev.} \textbf{D57} (1998) 1886;
  J.~Bl\"{u}mlein, \emph{Eur. Phys. J.} \textbf{C20} (2001) 683.

\bibitem{cafarella}
  A.~Cafarella and C.~Corian\`{o},
  \emph{Comp. Phys. Commun.} \textbf{160} (2004)~213.

\bibitem{shimizu}
  H.~Shimizu, G.~Sterman, W.~Vogelsang and H.~Yokoya,
  \emph{Phys. Rev.} \textbf{D71} (2005) 114007.

\bibitem{sps}
  M.J.~Corden \emph{et al.}, \emph{Phys. Lett.} \textbf{B68} (1977)~96;
  \emph{Phys. Lett.} \textbf{B96} (1980)~411;
  \emph{Phys. Lett.} \textbf{B98} (1981)~220.

\bibitem{carlson}
  C.E.~Carlson and R.~Suaya, \emph{Phys. Rev.} \textbf{D18} (1978)~760.

\end{thebibliography}
\end{document}